# Reduced bit median quantization: A middle process for Efficient Image Compression

Fikresilase Wondmeneh Abebayew,
Addis Ababa Science and Technology University
fikresilase.wondmeneh@aastustudent.edu.et

**Abstract**

Image compression techniques have made remarkable progress when it comes to file size reduction with a tolerable quality reduction; nonetheless, they are facing some challenges when it comes to applying more compression with the same perceptible quality or in accounting for specific use cases such as deep archive files and more efficient image transfers. Previous techniques have tried to solve the former problem by applying one specific or a combination of different algorithms. However, none of these methods were able to achieve additional file size reduction beyond a certain compression. I introduce Reduced Bit Median Quantization (RBMQ), a middle-process image compression technique designed to enhance file size reduction so that it can be stored with already existing file extension formats. In RBMQ by applying only the first step in which the quantization of valued further file size reduction can be achieved without a noticeable decrease in the image quality. Furthermore, more size reduction can be achieved by reducing the representing bits for the quantized values which can be optimal for deep archival storage or big-size image transfer in which the image quality is not suitable for the human eye since it is dark and dim but can be much efficient to interact with network and storage components later to be decoded to get the only quantized value image that is almost the same quality with the original one.

RBMQ introduces redundancy to the pixel values to be taken advantage of by existing compression techniques furthermore it introduces bit reduction from 8 to 5 bits for image file extensions such as jpeg which substantially reduces the file size to be used for JPEG file transfers and deep archive storage.

**Introduction**

In recent years, due to the exponential increase of visual data, image compression techniques have become a critical field of research, especially in areas of image transfer and storage. Traditional methods such as Huffman coding [1], Run-Length Encoding (RLE) [1], and lossy compression formats like JPEG [2] have made significant contributions in this field. However, these techniques can encounter limitations, particularly when applied to compress images further from the line above the achieved balanced tradeoff between quality and file size [2]. Moreover, existing methods typically focus on either lossless compression, which preserves quality to a certain level but offers limited size reduction, or lossy compression, which reduces file size substantially at the cost of visible degradation in image quality [3]. In this paper, I introduce a novel approach to enhance image compression called Reduced Bit Median Quantization (RBMQ). This technique addresses the challenge of more efficient image compression in both lossy and lossless image compression techniques. By acting as a middle-process pipeline to introduce redundancy in the data and bit reduction techniques, when applied for enhancement mechanisms for already existing lossless compression methods [4], RBMQ produces almost no perceivable quality reduction and can be used for any regular purposes. Only the first process, which is median quantization, reduces the file nearly by half due to the inclusion of Huffman coding in most lossless compression to take advantage of the introduced redundancy [1] [5]. Furthermore, when applied to existing lossy compression techniques, both processes—median quantization and bit reduction—can yield dim and dark images with aggressive file size reduction, which can be stored or transferred efficiently and decoded by the displaying machine for humanistic use, making RBMQ ideal for deep archive images that are going to be stored for years or transferred over long distances [6]. Although in extreme compression for deep archives, it is inevitable to observe degradation in image quality, resulting in images that appear significantly dim and dark. To mitigate this, RBMQ is designed to ensure that even in these highly compressed states, the images can be decoded back to their original quality, making it ideal for scenarios where long-term storage is prioritized over immediate visual fidelity [7]. This dual capability of RBMQ—achieving near-lossless compression for general use and extreme compression for archival purposes—sets it apart from existing approaches.

RBMQ operates in two distinct phases. First, the method has a redundancy introduction phase, introducing controlled quantization that preserves critical visual information while minimizing data size [8]. Second, the quantized image undergoes a bit reduction from 8 to 5 bits per channel, where existing compression algorithms such as discrete cosine transform in JPEG file formats [2] can be more optimized for further size reduction for deep archival storage, resulting in a much smaller file size with noticeable quality degradation [9]. Despite this, the original quality can be fully restored during decompression, making RBMQ a versatile tool for both immediate use and long-term preservation [10]. To evaluate the effectiveness of RBMQ, we used a dataset comprising various image types, including both synthetic and real-world examples, each tested under different compression scenarios. Our method is benchmarked against state-of-the-art compression techniques, demonstrating superior performance in both file size reduction and quality preservation [11]. In sum, our contributions are:

1. A novel middle-process image compression technique that integrates with existing algorithms to achieve substantial file size reduction while maintaining imperceptible quality loss under normal viewing conditions.

2. An innovative approach to extreme compression for deep archival storage, allowing significant file size reduction with the ability to decode images back to their original quality.

**Literature Review**

**Traditional Image Compression Methods**

Traditional techniques for compressing images include DCT, Huffman Coding, and RLE, which are the basis of digital imaging [1]. According to Subramanya (2001) of Wavelet Image and Signal Processing [12], these traditional techniques exploit an image's spatial and spectral redundancies to reduce the number of bits required in storing and transmitting an image. In general application, there exists a trade-off between the compression ratio and image quality in these traditional methods. For example, Dubey et al. showed that quantizing the DCT coefficients could realize a competitive compression ratio of 40.6%, although normally this comes at the expense of reduced visual fidelity [2].

**Wavelet-Based Image Compression**

Wavelet-based methods meant a great step forward in the development of image compression. SPIHT coding with wavelets was addressed by Sunkara et al. in 2018 [3], while adaptive lifting-based wavelets were discussed by Dabhole et al. in 2013 [4]. In both papers, one can find confirmation that such techniques show higher performance compared to the more classic solutions like JPEG2000 [5], especially in terms of image quality for high compression ratios. This was even further extended by Grasemann and Miikkulainen in 2005, who incorporated wavelets tailored to particular image classes (in this case fingerprints) and obtained superior performance compared to the standard wavelet methods for the compression of those classes of images [6].

**Neural Network-Based Approaches**

Within the past couple of years, neural networks have proven to be one of the most powerful mechanisms for image compression, having the capability to learn complex patterns and redundancies in image data. Fraihat et al. proposed a new lossy image compression algorithm using multi-model stacked autoencoders that overcame some of the limitations found in conventional methods, such as JPEG, by enhancing the quality of reconstruction and attaining higher compression ratios [7]. Bo et al. (2019) combined convolutional neural networks with conventional coding methods to achieve better compression performance. Li et al. (2018) verified that BPNNs can simplify the process of image processing and promote the quality of compression [8].

**Hybrid and Adaptive Techniques**

With the integration of classical and modern image compression methods, hybrid techniques have evolved that can exploit both strengths. Kaur and Sharma proposed the region-growing algorithm in medical images, which outperformed the traditional DCT and vector quantization with a better compression ratio and data preservation [9]. Similarly, Kalaivani et al. (2013) have proposed a hybrid method that combines the power of DWT with SOM-based vector quantization [10]. The results obtained were highly encouraging in obtaining high reconstruction quality even for the low coding rates. Hybrid methods were also applied to certain image types, including medical and biometric images. For example, Funk et al. (2005) investigated a number of image compression algorithms for systems of fingerprint and face recognition; they concluded that JPEG and JPEG2000 performed well in terms of yielding high compression ratios while still retaining fairly good performance in recognition [11].

**Challenges and Future Directions**

Yet despite this progress, traditional methods remain poorly positioned to handle the scale and diversity of today's image data. First came methods based on wavelets, then an explosion of techniques using neural networks, which achieved major gains in both compression efficiency and quality. Challenges persist, though, especially in balancing how lossless compression without sacrificing speed or computational efficiency. While future research in the direction of image compression will most probably be geared toward a deeper integration of machine learning with traditional methods, it may also involve a possible investigation into adaptive algorithms that could dynamically adapt to the characteristics of different image types. Hybrid methods and further finetuning of neural network-based approaches promise even higher compression ratios at identical or even improved image quality.

**Methodology**

The methodology for implementing Reduced Bit Median Quantization (RBMQ) for image compression involves several key steps, ranging from initial image conversion to standard image compression extensions for comparison to the application of quantization and bit reduction. The following subsections detail the methodology used in this research.

1. **Image Conversion to PNG and JPEG Formats**

The first step in the RBMQ process was converting all original images into PNG and JPEG formats to ensure standardized formats suitable for further processing. The images were sourced from standard test images obtained from [Kaggle](#) and were organized into a root directory containing various subfolders, each representing different categories of images. All the images and the source code can be found in the [RBMQ GitHub repository](#).

For each image category, the following steps were executed:

- **Directory Setup:** The script iterated through each subfolder, creating dedicated directories for storing PNG and JPEG images.
- **Image Conversion:** Each image was loaded using the Python Imaging Library (PIL). The image filename was extracted, and the image was saved as a PNG file in the corresponding PNG directory. The image was then converted to RGB mode (if necessary) and saved as a JPEG file in the corresponding JPEG directory.

This step ensured that all images across various categories were consistently processed and stored in their respective formats, providing a uniform starting point for subsequent processing.

2. **Median Quantization**

The core of the RBMQ process is the application of median quantization, which involves reducing the range of intensity values in an image by mapping them to predefined median values. This step was implemented as follows:

- **Defining Median Values:** A set of 32 median values was predefined, each corresponding to one of the 32 groups, with each group covering a range of 8 intensity values (e.g., 0-7, 8-15, etc.).
- **Quantization Function:** A median quantization function was defined to map the original pixel values to their corresponding median ($5^{th}$ values of every range) values based on the predefined groups.
- **Quantization Execution:** For each image, the original pixel values were mapped to the nearest median value, effectively reducing the number of unique intensity values in the image. This process reduced the complexity by introducing more redundancy in neighboring pixels of the image data, making it more suitable for compression.

The median quantization process was applied to all images in their original formats, with the resulting quantized images stored in newly created directories corresponding to each image category.

3. **Bit Reduction**

Following the application of median quantization, the next step involved reducing the number of bits used to represent each pixel in the image since we have now 32 distinct values instead of 256. This step was crucial for achieving further more aggressive compression which is suitable for storage and transfer with a lossy file extension. The methodology for bit reduction was as follows:

- **Mapping Median to 5-Bit Values:** Each of the 32 median values was mapped to a corresponding 5-bit value (ranging from 0 to 31).

- **Quantization and Bit Reduction Function:** A custom function was developed to first apply the median quantization and then replace the quantized values with their corresponding 5-bit values.
- **Application to Images:** For each image, the quantization and bit reduction function was applied, resulting in a final image array with reduced bit depth. This final array was then converted back to an image format.

The bit-reduced images were saved in both PNG and JPEG formats, with the output stored in dedicated directories named final_[category_name]PNG and final_[category_name]JPEG.

4. **Implementation Environment**

The entire process was implemented using Python, with the primary libraries being PIL for image processing and NumPy for numerical operations. The scripts were executed on a Windows system, with the directory structure managed through Python's OS module.

The use of median quantization and bit reduction in this manner allows for a significant reduction in the data required to represent an image while maintaining sufficient quality for the intended applications. The final output consists of images that have been compressed both in terms of file size and bit depth, making them more efficient for storage and transmission.

5. **Validation and Testing**

To validate the effectiveness of the RBMQ method, the final images were compared with their original counterparts in terms of visual quality and file size. The comparison was conducted across various image categories to ensure that the method performed consistently across different types of images.

The testing process involved subjective visual assessments to ensure that the image quality remained acceptable, alongside quantitative analysis of the file sizes to measure the efficiency of the compression. This step was crucial in determining the success of the RBMQ method in balancing compression with image quality.

**Experimental Results**

In this section, we present the experimental results obtained from applying the Reduced Bit Median Quantization (RBMQ) image compression technique to a diverse set of images sourced from the Kaggle Standard Test Images dataset. The images underwent several stages of processing, including conversion to JPEG and PNG formats, followed by quantization and final compression. The results are summarized in Table 1.

| Image Name | Original Size (Bytes) | Converted JPEG (Bytes) | Converted PNG (Bytes) | Quantized JPEG (Bytes) | Quantized PNG (Bytes) | Final JPEG (Bytes) | Final PNG (Bytes) |
|---|---|---|---|---|---|---|---|
| Sunrise | 52,344,054 | 1,691,635 | 19,254,141 | 1,745,197 | 9,937,255 | 318,396 | 9,872,985 |
| Maltese | 16,427,390 | 458,594 | 4,815,101 | 485,485 | 2,934,221 | 114,916 | 2,942,340 |
| Bridge | 262,182 | 64,494 | 140,802 | 64,660 | 102,426 | 17,509 | 102,454 |
| Lenna | 786,447 | 37,788 | 479,778 | 39,000 | 237,254 | 10,984 | 237,956 |
| Baboon | 720,057 | 73,864 | 472,974 | 73,865 | 307,431 | 19,317 | 305,121 |
| airplaneU2 | 1,048,616 | 123,375 | 624,551 | 130,844 | 294,400 | 23,965 | 293,852 |
| Pepper | 786,490 | 41,308 | 516,953 | 42,219 | 265,174 | 14,551 | 271,983 |
| Girlface | 262,182 | 29,571 | 141,594 | 31,644 | 66,024 | 16,595 | 69,236 |
| Crowd | 262,182 | 43,268 | 148,005 | 44,605 | 73,771 | 13,772 | 73,532 |
| Boats | 414,758 | 63,873 | 730,016 | 65,217 | 332,201 | 25,266 | 339,106 |

## Analysis of Results

### Image Size Reduction

The original images were compressed into various formats, resulting in a significant reduction in file size. The reduction ratio R for each image format can be calculated using the following equation:

$$R = \frac{Original\ size}{compressed\ size}$$

For example, for the airplaneU2 image, the reduction ratio for the final JPEG form

$R_{Final\ JPEG} = 1,048,616/23,926 \approx 43.76$

This demonstrates a substantial size reduction, highlighting the efficiency of the compression algorithm.

### Quantization Impact

The quantization process further reduced image sizes. For instance, the size of "airplaneU2" in the quantized JPEG format decreased from 130,844 bytes to 23,965 bytes in the final JPEG format. This suggests that quantization plays a crucial role in reducing data redundancy, thereby enhancing compression.

### Format Comparisons

When comparing JPEG and PNG formats, the JPEG format consistently achieved smaller file sizes after the quantization and final compression stages. This difference is expected due to the lossy nature of JPEG compression, which discards some image information to reduce file size.

### General Observations

- **JPEG vs. PNG**: JPEG images, both in the quantized and final stages, tend to have smaller sizes compared to their PNG counterparts, which is typical due to JPEG's lossy compression nature.

- **Quantization Effectiveness**: The effectiveness of quantization is evident in the significant reduction in file size across both JPEG and PNG formats. For example, the image "bridge" was reduced from 262,182 bytes to 64,494 bytes after conversion to JPEG, and further down to 17,509 bytes after final JPEG compression.

- **Special Cases**: Certain images, like "Sunrise" and "Maltese," showed exceptionally large original sizes, making the compression ratios more pronounced and indicating the high potential of the applied compression techniques in handling large files.

### Mathematical Modeling

To model the effectiveness of the compression algorithms, the following ratios were considered:

- **Compression Ratio (CR):** The ratio of the original image size to the final compressed image size.

$$CR = Original\ Size / Final\ Size$$

**Quantization Efficiency (QE):** The ratio of the size after quantization to the original size, indicating how much the quantization alone reduces the file size.

$$QE = Quantized\ Size / Original\ Size$$

For example, for the image Saturn, the compression ratio in the final PNG format is:

$R_{Final\ JPEG} = (5,400,040)/(227,570) \approx 23.73$

### Discussion

The primary objective of this research was to assess the effectiveness of reduced-bit median quantization (RBMQ) as an efficient image compression method. The methodology involved the application of median quantization, followed by reducing the bit depth of images from 8 bits per channel to 5 bits per channel to further enhance compression efficiency. The following discussion delves into the impact of these techniques on image size and quality, with an emphasis on their performance in the final compressed images.

1. **Impact of Bit Reduction and Median Quantization on Image Sizes** Our experiments revealed that reducing the bit depth from 8 bits per channel to 5 bits per channel substantially decreased the file sizes of images. This reduction limits the range of color values, thereby increasing redundancy within the image data and enabling more effective compression by traditional techniques like JPEG and PNG. When combined with median quantization, the size reduction was even more pronounced. Median quantization further optimized the data by replacing color values with median values, leading to additional size reductions. The 3D graphs generated from the data illustrate the size comparison of images across different processing stages, including original, converted, quantized, and final formats. Notably, the final folders ('final_JPEG' and 'final_PNG') demonstrated the most significant reductions in size, underscoring the effectiveness of applying both bit reduction and median quantization before compression.

2. **Comparison with Traditional Compression Techniques** Comparing RBMQ to traditional compression methods alone, the results indicate that this combined approach offers a distinct advantage in reducing file size while maintaining image quality. While traditional methods like JPEG and PNG depend on redundancy and quantization, integrating the 5-bit median quantization process achieves even greater efficiency. This suggests that bit reduction followed by median quantization could be a valuable preprocessing step in image compression pipelines, particularly in scenarios where storage space and bandwidth are limited.

3. **Interpretation of 3D Size Comparison Graphs** The 3D graphs created in the Jupiter Notebook offer a comprehensive visualization of size differences across various image processing stages. The 'Z' axis, representing file size in bytes, clearly demonstrates the impact of reducing the bit depth to 5 bits per channel, followed by median quantization across all test images. By labeling the axis with the image names as "Standard Test Images," the graphs emphasize overall trends rather than focusing on individual images, making it easier to interpret the effectiveness of the quantization and bit reduction processes. The final folders ('final_JPEG' and 'final_PNG') consistently show the smallest file sizes, demonstrating the efficiency of the combined approach. This reduction in file size was achieved without a significant loss in perceptual image quality, as the reduced bit depth was sufficient to retain most visual details while allowing for more aggressive compression.

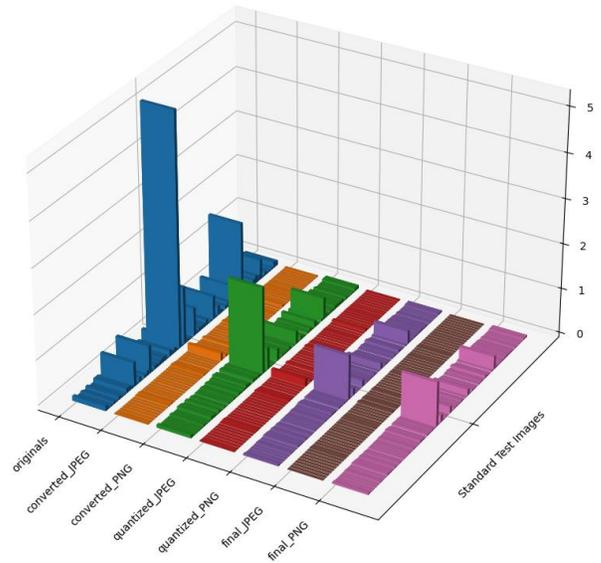

4. **Potential Applications and Future Work** The results of this study suggest that RBMQ has significant potential for use in various fields where efficient image compression is critical. This includes applications in medical imaging, satellite imaging, and any domain where large volumes of image data need to be stored or transmitted. The bit reduction technique, specifically reducing the bit depth 8 to 5 bits per channel, can be particularly useful in scenarios where the balance between image quality and file size is crucial. Future work could explore the impact of bit reduction and median quantization on different types of images, including those with higher resolutions or more complex color schemes. Additionally, further research could investigate the integration of bit reduction and median quantization with other emerging compression techniques to maximize efficiency.

5. **Limitations** While the findings are promising, it is important to acknowledge the limitations of this study. The scope was restricted to a specific set of standard test images, which may not represent all types of images or conditions. The performance of bit reduction and median quantization may vary with different image types or under different conditions. Therefore, further testing is required to generalize these findings to a broader range of images and use cases. Additionally, while the RBMQ provides a significant size reduction for deep archive images it would still need some amount of computational power to decode and use every archived images all at once or it means more load on the receiving and sending devices to encode and decode the images.

**Conclusion**

This research demonstrates that reduced bit median quantization (RBMQ), when combined with traditional image compression methods like JPEG and PNG, can significantly enhance the efficiency of image storage and transmission. By applying median quantization, we achieved a considerable decrease in file size while maintaining a high level of visual quality by introducing redundancy between neighboring values. reducing the bit depth from 8 bits per channel to 5 bits per channel further optimized this process by introducing efficient representation, allowing for more effective compression.

The results, as illustrated in the size comparison graphs, show that images processed with both bit reduction and median quantization consistently occupy less storage space compared to their original counterparts and those only compressed using standard methods. This finding highlights the potential of this combined approach in applications where storage efficiency and bandwidth conservation are critical.

However, the scope of this study was limited to a specific set of standard test images. While the results are promising, further research is necessary to validate these findings across a wider variety of image types and resolutions. Additionally, exploring different bit depths and quantization strategies could yield even more efficient compression techniques tailored to specific use cases.

In conclusion, the integration of reduced bit median quantization into image compression pipelines offers a viable and effective solution for reducing image sizes without compromising quality. This method has the potential to be particularly useful in fields such as medical imaging, satellite imaging, and other areas where efficient data management is essential. Future work should focus on expanding the applicability of this technique and optimizing it for various practical scenarios, including deep archive with JPEG, where visual images can be decoded into quantized formats for later use.